\documentclass[12pt]{article}
\usepackage {graphics}
\begin{document}
\noindent
{\Large Extracting useful information from noisy exponentially decaying signals }

\vspace{2cm}

\noindent
{\large David Lawunmi}

\noindent
Department of Oncology 

\noindent
Royal Free and University College Medical School 

\noindent
University College London

\noindent
Royal Free Campus

\noindent
Rowland Hill Street 

\noindent
London NW3 2PF

\vspace{4cm}

\noindent
Analysing data that consists of one or more truncated exponential functions is of great interest in a wide range of fields. 
Data consisting of one or more exponential functions are measured by a wide range of instruments, examples include: nuclear magnetic 
resonance, (NMR); magnetic resonance imaging (MRI); the analysis of radioactive decay data; chromatography; lifetime fluorescence imaging;
 spin resonance, (ESR). Many of the algorithms that are currently in used for  \textquoteleft characterising\textquoteright \hspace{0.2cm}
this data, are difficult to interpret or are they useless and they produce erroneous results. Some of the key technical issues associated
with this problem are discussed in this article we also present some results that use some elementary properties of exponential properties
of exponential and harmonic functions. These properties allow us to develop relationships between these functions which we exploit to
analyse exponentially decaying signals in noisy environments. 

\pagebreak

\medskip
\noindent
\begin{center}{\Large\textbf{Introduction}}\\
\end{center}

\noindent
In this article we derive a simple technique for analysing and characterising data consisting of one or more truncated 
exponential functions.
We exploit a number of simple mathematical relationships between these functions and the harmonic functions, 
$\cos(x)$, and $\sin(x)$. We use these relationships to derive
expressions that enable us to characterise truncated exponential signals from single exponential functions in noisy environments. This
approach can be generalised to
characterise data from signals consisting of more than one exponential function in a noisy environment.
Many physical phenomena are described by systems of one or more differential equations. The solutions of these systems will in general
consist of one or more  exponentially decaying functions. In order to use the data generated by these systems to determine their physical 
characteristics it would be useful to have a robust mathematical technique that could determine the decay constants and amplitudes of 
each of the exponentially decaying component signals that occur in a set of measurements. In particular it would be useful to be able 
characterise truncated exponential signals.

\medskip
\noindent
Signals consisting of one or more exponential functions arise in a number of areas, examples include: deep level transient 
spectroscopy, in semiconductor physics, Istratov and Vyvenko (1999); fluorescence decay analysis in the analysis of biological samples 
Jones et al (1999); the analysis and characterisation of time series measurements from radioactive samples, Cottingham and 
Greenwood, (2001); the analysis of the relaxation phenomena that occur in nuclear magnetic resonance studies of materials, Cowan, 
(1997); reaction kinetics, \'{E}rdi and T\'{o}th, (1989); positron emission tomography, Valk et al (2003); MRI, Wells (1982), 
the Biacore, Fersht (1999).

\medskip
\noindent
Attempts at doing a least squared fit of one or more exponentials to data and varying the decay constant(s) and the amplitude(s)
to get a best fitting exponential function are fraught with algorithmic difficulties. The results are often misleading and ambiguous,
even when the noise level in the data is low. These approaches are still unfortunately  in wide use. Some authors seem to 
be unaware of the technical problems that are associated with processing data that consists of one or more exponential functions,
e.g. Baxter et al (1994) and (1995), Lubic (2001), Valk (2003).
The algorithmic problems associated with attempting to fit exponential functions to data have been discussed by a number of authors 
e.g. Sivia (1996), Gans (1992), Istratov and Vyvenko (1999), Fersht (1999), Acton (1990).

\medskip
\noindent
The approach that we are suggesting in this article should help to provide a mathematical basis for choosing the points in time at which
data can be collected; should provide a means of analysing and characterising truncated exponential signals, this is a major technical
problem as people that analyse data that contains even a single exponential tend to wait till the signal is close to its zero level before
attempting to characterise the signal, Cowan (1997); the technique also provides a link between harmonic functions and exponential 
functions, that can still be of value when noise is present; in principle it should be possible to generalise this work in order to  
analyse data consisting of more than one exponential function. Making the link between exponential and harmonic functions may enable us 
to develop ways of improving on the resolution arguments that were given by Istratov and Vyvenko, (1999), for analysing data consisting of
a number of exponentially decaying signals.

\medskip
\noindent
\begin{center}{\Large\textbf{Mathematical Analysis}}\\
\end{center}

\medskip
\noindent
The techniques that we develop in this article are concerned with the analysis and characterisation of data of the form

\begin{equation}\label{lab_0}
f(t) = \sum^{\infty}_{j=1} C_{j}e^{-\lambda_j t} + n(t)
\end{equation}

\noindent
The noise is expressed by the term n(t). 
This term or function can be thought of as a stochastic function of the time. 
We begin the analysis by utilising some properties of Chebyshev polynomials.

\noindent
The Chebyshev expansion of an exponential function has the form

\begin{equation}\label{lab_1}
e^{- \alpha x} =   2I_0(\alpha) T_0(\alpha x) +2 \sum_{j=1}^{\infty} T_j(\alpha_k x) I_j(\alpha) (-1)^j
\end{equation}

\medskip
\noindent
Arfken and Weber, (2000)
where $I_{m}$ represents modified Bessels functions of order m. The Chebyshev expansion occurs over an interval, $-1 \leq x \leq 1$.
In practice when data has been collected over an interval such that $x_{min}$ $\leq x \leq x_{max}$, this interval can be converted to the
interval $-1 \leq \overline{x} \leq 1$, by using the transformation

\begin{equation}\label{lab_2}
\overline{x} = \frac{2x - x_{min}-x_{max}}{x_{max}-x_{min}}
\end{equation}

\noindent
Thus the Chebyshev expansion for an exponential function can be generalised to an exponential function that is defined over a general 
finite interval, e.g. $-1 \leq x \leq 1$. 

\noindent
Consider a  signal, $f(t)$, consisting of an exponential function and some noise, where

\begin{equation}\label{lab_2aa}
f(t) = C e^{- \lambda t} + n(t)
\end{equation}

\noindent
$f(t)$, can be expressed as a series of Chebyshev polynomials, where

\begin{equation}\label{lab_3}
f(t) = \sum^{\infty}_{n=0}A_{n}T_{n}(x)
\end{equation}

\noindent
In practice the sum will of course be done over a finite number of terms, and so the function f(t) will be represented by an
approximation of the form

\begin{equation}\label{lab_4}
f(t) \approx  \sum^{N_{max}}_{n=0}A_{n}T_{n}(x)
\end{equation}

\noindent
where $N_{max}$ is the highest order term used in the Chebyshev expansion for $f(t)$.
This process facilitates the reduction of noise. Noise reduction is achieved by expanding the data in terms of
Chebyshev polynomials and discarding all of the terms above a particular order. The value of $N_{max}$ can be estimated or decided upon 
by the user.  For example this could be acheived by prior experimentation on a time series containing exponential functions and some 
computer generated pseudo random noise with the characteristics that are likely to be generated by the equipment that measures the signal
of interest.

\medskip
\noindent
Relating a series of exponential functions in a noisy environment to a series of harmonic functions in a noisy environment can be 
acheived by  converting a series of Chebyshev polynomials, $T_{n}(x)$, to a power series in x. Chebyshev polynomials can be defined by the
the following expressions

\begin{equation}\label{lab_5}
T_{0}(x) = 1
\end{equation}

\begin{equation}\label{lab_6}
T_{1}(x) = x
\end{equation}

\noindent
and the recursion relation

\begin{equation}\label{lab_7}
T_{n+1}(x)= 2xT_n(x) - T_{n-1}(x)
\end{equation}

\noindent
Fox and Parker, (1968).

\noindent
Consider an exponential function, this can be expressed in terms of a power series 

\begin{equation}\label{lab_8}
e^{ax} = \sum^{\infty}_{n=0} \frac{(ax)^{n}}{n!}
\end{equation}

\noindent
This can of course be approximated by

\begin{equation}\label{lab_9}
e^{ax} \approx \sum^{N_{max}}_{n=0} \frac{(ax)^{n}}{n!}
\end{equation}

\noindent
where the $N_{max}$th order term i.e. $x^{N_{max}}$, is the highest order term that is used in the approximation.
The expansion for  $\sin (ax)$, is given by

\begin{equation}\label{lab_10}
\sin{ax} = \sum^{\infty}_{n=0} \frac{(-1)^{n}(ax)^{2n+1}}{(2n+1)!}
\end{equation}

\noindent
This can be approximated by

\begin{equation}\label{lab_11}
\sin{ax} \approx \sum^{N_{max}}_{n=0} \frac{(-1)^{n}(ax)^{2n+1}}{(2n+1)!}
\end{equation}

\noindent
similarly the expansion for $\cos(ax)$ is given by

\begin{equation}\label{lab_12}
\cos{ax} = \sum^{\infty}_{n=0} \frac{(-1)^{n}(ax)^{2n}}{(2n)!}
\end{equation}

\noindent
The algorithm involves expanding the exponential in terms of some polynomials or splines, e.g. Chebyshev polynomials. The odd order terms
are separated from the even order terms. The odd Chebyshev polynomials are then expanded out into a set of terms that  consist of powers 
of x. 

\noindent
Similarly the even terms can also be expanded into a power series. One way of converting the power series of an exponentially decaying
signal where noise is not present to a power series for cosine and sine functions is to multiply all of the negative terms in the power 
series that is obtained for $e^{-ax}$ by $-1$, i.e. terms of the form $x^{r}$, are multiplied by -1. This yields the power series for,
$e^{ax}$. Now consider the odd order terms in the power series, terms where the power in, $x^{r}$, satisfies, $r = 2q	+ 1$.

\noindent
If $\frac{r-1}{2}$, is even or equal to zero, the term is left untouched. If $\frac{r-1}{2}$, is odd the term is multiplied by -1.
This yields the power series for $\sin(ax)$.

\noindent
Consider the even order terms in the power series. If $\frac{r}{2}$, is even or equal to zero, the term is left untouched. 
If $\frac{r}{2}$, is odd the term is multiplied by -1. This yields the power series for $\cos(ax)$.

\noindent
Thus the even powers of the expansion of the function, $f(x) = e^{ax}$, when processed as above will yield the function
$\cos(ax)$. 

\medskip
\noindent
This process can be thought of in terms of obtaining a power series in $x$ and then replacing $x$ by the imaginary
number $z=ix$, where $i = \sqrt{-1}$. Thus 

\begin{equation}\label{lab_14aaa}
e^{x} \rightarrow e^{ix} \equiv \cos(x) + i \sin(x)
\end{equation}

\noindent
The harmonic functions that are generated from the exponenentially decaying (growing) signals can of course be processed by a range of
standard techniques such as Fourier based techniques, such as the fast Fourier transform. In practice it is likely that the user will 
only have a small section of the harmonic signal, that spans less than a full periodic cycle of the harmonic functions that are generated 
by the above process. One way of  overcoming this is to multiply the expansions for $\cos(ax)$ and $\sin(ax)$by a suitable harmonic 
function and then to recombine them. Consider the following trigonometric relationship

\begin{equation}\label{lab_14}
s(x) = \cos(\alpha x) \cos(\Omega x) - \sin(\alpha x)\sin(\Omega x) \equiv \cos(\Omega x + \alpha x)
\end{equation}

\noindent
This is generated by multiplying the term $\cos(\alpha x)$, by $\cos(\Omega x)$, and the term, $\sin(\alpha x)$, by $\sin(\Omega x)$, 
If $\Omega$, is chosen so that it varies by at least one cycle over the interval in which the power series approximations for 
$\sin(\alpha x)$ and $\cos(\alpha x)$ were generated, it should result in there being more than one cycle of the function 
s(x) over the interval of the variable x  in which we have generated approximate power series for, $e^{\alpha x}$,  $\cos(\alpha x)$
and  $\sin(\alpha x)$. A number of other harmonic functions can be generated from the power series for $\cos(\alpha x)
$ and $\sin(\alpha x)$, with the aid of the standard trigonometric identities, e.g.

\begin{equation}\label{lab_15}
\sin([\Omega \pm \alpha]x) \equiv \sin(\Omega x)\cos(\alpha x) \pm \cos(\Omega)\sin(\alpha x)
\end{equation}
\noindent
and

\begin{equation}\label{lab_16}
\cos([\Omega \mp \alpha]x) \equiv \cos(\Omega x)\cos(\alpha x) \pm \sin(\Omega)\sin(\alpha x)
\end{equation}

\noindent
If $\Omega$ is chosen appropriately this process results in the generation of one or more cycles of a harmonic function.

The functions $\sin([\Omega \pm \alpha]x)$ and  $\cos([\Omega \pm \alpha]x)$can then be analysed by using a suitable harmonic analysis 
technique e.g. the periodogram or a fast Fourier transform.  If data is collected over a number of closely spaced data points this will 
tend to benefit the harmonic frequency analysis of the periodic function that is generated from the exponential. 
This process can be generalised to a range of data intervals, e.g. in practice a user may not wish to set the origin in the middle of some
interval of the dependent variable. It may be more advantageous to obtain an approximation for the exponential function and subsequently 
the sine and cosine functions by expanding a point other than $t = 0$.

\begin{center}{\Large\textbf{Some results}}\\
\end{center}

\noindent
Consider a signal that consists of a single exponential function, $e^{-\lambda x}$, where the noise term is equal to zero. If the data is 
collected between the points $t = 0$ and $t=t_{max}$ , this can be rescaled to $\frac{-t_{max}}{2} \le t^{\prime} \le \frac{t_{max}}{2}$, 
using equation (\ref{lab_2}), we can define a new variable $\overline{x}$, where $-1 \leq \overline{x} \leq 1$. The data will be 
collected at a finite number of data points in this interval, note the points do not need to be evenly spaced. 
A Chebyshev expansion can be obtained for this signal using software such as the NAG subroutine E02ADF. The Chebyshev expansion can be use
to generate a power series 
expansion in terms of the variable $\overline{x}$. From the arguments presented above this can be converted into two power series, one for
$\sin{(\lambda x)}$, and one for $\cos(\lambda x)$. The cosine series is generated from the even order terms, i.e. terms of order,
$0, 2, 4, \cdots$, and the sine series from the order terms, i.e. terms of order, $1, 3, 5, \cdots$. The above approach works well in the
zero noise case. Some results for the noise free case are presented in figure (1a), figure (1b), figure (1c) and figure (1d). 
In figure (1a) a plot of an exponential composed of twenty five equally spaced data points is presented. The corresponding sinsuoid and 
cosinusoid are presented in figures (1b) and (1c). In figure (1d) a plot based on expression (\ref{lab_14}) is presented, the angular 
frequency corresponding to $\Omega  + \alpha $, is 23.56 $s^{-1}$. In these figures, $t_{min} \leq t \leq t_{max}$, where, $t_{max} = 1s$,
$t_{min} = -1s$, and thus the variable $\overline{x}$, is equivalent to the variable, $t$.

\medskip
\noindent
Some plots for a simulated data set consisting of 250 evenly spaced and normally distribiuted noise with a standard deviation of:
1 $\times$ $10^{-2}$ units, 1 $\times$ $10^{-1}$ units, and 5 $\times$ $10^{-1}$ units, are presented in figures (2), (3) and (4)
respectively, in all of the cases the mean of the noise is 0.0 (units) and the angular frequency corresponding to $\Omega  + \alpha $, 
is 23.56 $s^{-1}$. These plots illustrate the possibility of extracting a harmonic signal from a noisy exponential even when the noise 
level is relatively high. Though the most accurate data will tend to be located around the zero point, i.e. the point $t = 0$, which was
used to generate the polynomial expansion. If the subsequent harmonic analysis concentrates on a few cycles of the harmonic signal 
generated about the point $t=0$, a few cycles of a harmonic signal that is very is close in value to the corresponding noise free 
harmonic function is obtained. We have investigated the effectiveness of the technique with a simulated exponential function 
using normally disctributed noise with a mean of 0.0 (units) and a range of values for the standard deviation. 
In figure (1a) 25 data points were used. 
In figures(3a), (4a) and (5a) 250 data points were used. In general for a given noise 
level the quality of the functions that are generated by the processes that we describe in this article tends to improve with the number
of data points of the input exponential function that are used to generate harmonic functions.

\begin{center}{\Large\textbf{Discussion}}\\
\end{center}
A simple technique that is based on properties connected to the power series of harmonic functions and exponential functions has been
developed. It can be used as a basis for relating an exponential function with a particular decay constant to an odd (sinusoidal) and an
even (cosinusoidal) function with a corresponding frequency of oscillation. In general it may be unrealistic to obtain more than a 
fraction of a cycle of these harmonic signals with this approach. In order to obtain more than a full cycle of a harmonic signal
we can avail ourselves of the standard trigonometric identities for the cosines and the sines of the sum of two angles (\ref{lab_14}) 
and (\ref{lab_15}). This enables us to combine the harmonic signal that was generated from the exponential signal with cosine and sine
functions with a frequency of oscillation that goes through one or more cycles over the time interval for which data was collected.
If the frequency of oscillation of the high frequency component is chosen appropriately this process should enable us to obtain several
cycles of oscillation of the harmonic signal that results from this process. In particular if the exponential signal is not noise free, 
this should enable us to obtain several cycles of harmonic signal that are not overwhelmed by the noise that was present in the exponentia
data. From our earlier analysis it appears that the harmonic component that is due to the exponential signal tends to be close in value 
to the harmonic signal that would be obtained from noise free data for points that are not too distant from the point about which the 
polynomial expansion was generated, ($t=0$), in this example. This is useful as it can be used as a basis for further analysis to
establish the frequency of oscillation of the resulting harmonic signal. We have thus demonstrated the feasibility of generating harmonic
signals from exponential signals. This process can be used to analyse and characterise small sections of truncated exponential signals, 
without having to wait for the signal to decay down to its tail region,  which is the convetional way of collecting data from a signal 
consisting of an exponential component and noise, e.g when analysing decay constants from an NMR analysis. 
If some information is available on the nature of the noise that is likely to be present with the desired signal, this process allows 
simulations to be performed to assess issues such as: the number of data points required; the spacing of data points; 
the quality of the signal over a particular time interval; the impact of noise on the exponential data on the harmonic signal(s) that are 
generated from this process; a false alarm analysis to estimate the probability of making a reasonable estimate of the decay constant for 
a particular noise type and noise intensity.

\medskip
\noindent
This approach can be generalised in a number of ways. We have focused on truncated exponential functions in noisy environments in this 
article. In principle this approach can be extended to look at signals with contributions from several exponential functions. This is
significant as it may enable us to improve on the resolution limits suggested by Istratov and Vyvenko (1999), for signals that contain
more than one exponential function.

\medskip
\noindent
We can exploit the properties of Chebyshev polynomials to integrate the time series. This allows us to obtain an analytic expression for 
the time series along with a constant of integration. Integration of a Chebyshev series can improve the signal to noise ratio of the 
Chebyshev expansion of the underlying functions, Fox and Parker (1968), Press et al (1992). Integration is also a useful means of changing
the relative intensities of the component exponential signals.

\medskip
\noindent
Another generalisation that is important to be aware of is that a signal that is composed of one or more exponential functions and noise
can be used to generate two signals, one composed of sinh functions and noise, and the other composed of cosh functions and 
noise. This may prove to be beneficial as these functions have useful mathematical identies that may be exploited when attempting to 
characterise the exponential functions in the data set, e.g.

\begin{equation}\label{lab_18}
\sinh([\Omega \pm \alpha]x) \equiv \sinh(\Omega x)\cosh(\alpha x) \pm \cosh(\Omega)\sinh(\alpha x)
\end{equation}

\begin{equation}\label{lab_19}
\cosh([\Omega \mp \alpha]x) \equiv \cosh(\Omega x)\cosh(\alpha x) \mp \sinh(\Omega)\sinh(\alpha x)
\end{equation}

\medskip
\noindent
They can be used as an alternative to generating the harmonic functions from the exponential functions. If this route is taken, the sinh
component generates the sinusoidal signals, and the cosh component generates the cosinusoidal component of the harmonic signal.

\medskip
\noindent
Use of polynomials, e.g. splines, and Chebyshev polynomials to approximate functions has certain advantages that prove to beneficial
for analysing certain types of signals. They generate an analytic approximation to the function of interest. This may be integrated,
and if appropriate it may be differentiated as well, (differentiation tends to be noisier than integration). These processes allow
the relative intensitites of component signals to be altered, this may help to find a component with a small amplitude, and a small
decay constant. It also facilitates the generation of signals with a large number of points. This does not violate information theory.
Using polynomial interpolation to estimate the value of the signal at points other than those at which it was estimated will of course
in general, produce an error at the points in time where information is required if a measurement was not made at the point(s) of 
interest. Thus a high density set of estimates of a quantity of interest, that was obtained from a sparse set of measurements will
tend to be much noisier than a set of real measurements of a signal that were made at all of the time points of interest.

\medskip
\noindent
Another point of interest with regards to integration is that harmonic signals can be generated by integration. For example, if the 
original signal had the form

\begin{equation}\label{lab_20}
s(t) = A \; \cos(\alpha x) + n(t)
\end{equation}

\noindent
Integrating this signal results in a signal of the form

\begin{equation}\label{lab_21}
s^{\prime}(t) = \frac{A\;\sin(\alpha x)}{\alpha} + n^{\prime}(t)
\end{equation}

\noindent
It may be possible to utilise one or both of expressions (\ref{lab_15}), and (\ref{lab_16})to generate a harmonic function that oscillates
over several cycles for a time interval over which the data has been collected. This signal can in principle be analysed by standard 
Fourier techniques. It has the form

\begin{equation}\label{lab_22}
s^{\prime \prime}(t) = A^{\prime \prime}\cos(\Omega x + \alpha x) + n^{\prime \prime}(t)
\end{equation}

\noindent
A similar approach will apply to the signal

\begin{equation}\label{lab_23}
\widehat{ s(t)} = A \; \sin(\alpha x) + n(t)
\end{equation}

\noindent
It is important to mention the Bayesian approach to analysing signals consisting of one or more exponential, most authors 
discuss the resolution constraints that are associated with this method, e.g. Ruanaidh and Fitzgerald, (1996). This approach is useful
as it comes with an in built statistical analysis as a means of quality control. In principle it should  be possible to combine a Bayesian
approach with the above approach to analyse how this technique responds to the type of noise that is expected to be associated with the
measurements, or alternaively to estimate the properties of the noise that are associated with a set of measurements. 

\medskip
\noindent
A simple alternative to this is to undertake a false alarm analysis by performing a large number of simulations on data that corresponds 
to the expected signal, and the noise that is expected to be associated with a set measurements. This should enable an estimate to be
made of the fitness for a particular purpose of this technique, for a data set of interest.

\bigskip
\begin{center}{\Large\textbf{Conclusion}}\\
\end{center}

\noindent
We have developed a simple technique that uses elementary mathematics, the power series of exponential, cosine and sine functions, standar
trigonometric identities and, and basic polynomial interpolation theory. This approach is simple to implement and it is also straight 
forward to test it for fitness for the intended purpose. We have also used elementary trigonometry to  derive  simple expressions
that may enable the analysis and characterisation of a fraction of a sinusoidal function. This is potentially of great value, in
particular when dealing with harmonic functions with a small circular frequency relative to the time over which it is possible to
collect the data that generates the harmonic functions.
It follows this work may be of value if it is necessary to analyse a data set that may contain a truncated exponential function. This 
facility allows the user to concentrate on the high signal to noise end of the data set and it may also result in a significant time
saving if the decay constant is small. This work also has the potential to be extended so that it can be applied to analyse data sets that
consist of more than one exponential function. This may result in a simple approach to analysing a problem that arises in a wide range
of experimental techniques.

\pagebreak
{\Large References}

\medskip
\noindent
\begin{enumerate}

\item J. K. \`O. Ruanaidh and Fitzgerald, Numerical Bayesian Methods Applied to Signal Processing, Springer Verlag, (1996)

\item  W. H. Press, B. P. Flannery, S. A. Teukolsky, and W. T. Vetterling, Numerical Recipes in Fortran, Cambridge University Press, 
(1992)

\item L. Fox and I. B. Parker Chebyshev polynomials in numerical analysis,  Oxford University Press, (1968)

\item A. Fersht, Structure and Mechanism in Protein Science: A Guide to Enzyme Catalysis and Protein Folding
W. H. Freeman \& Company, (1999) 

\item R. Jones,  K. Dowling,  M. J. Cole,  D. Parsons-Karavassilis, M. J. Lever, P. M. W. French, J. D. Hares, A. K. L. Dymoke-Bradshaw,
Electronics Letters, $\underline{35}$, p256, (1999)

\item W. N. Cottingham and  D. A. Greenwood, An Introduction to Nuclear Physics, Cambridge University Press,  (2001)

\item A. A. Istratov and O. F. Vyvenko, Review of Scientific Instruments, p. 1233, $\underline{70}$, (1999)

\item B. Cowan, Nuclear Magnetic Resonance and Relaxation, Cambridge University Press, (1997) 

\item P. \'{E}rdi and J. T\'{o}th, Mathematical models of chemical reactions : theory and applications of deterministic and 
stochastic models, Manchester University Press, (1989)

\item P. E. Valk (Editor), D. L. Bailey, D. W. Townsend, M. N. Maisey, Positron Emission Tomography: Principles and Practice, (2003)

\item P. N. T. Wells, Scientific Basis of Medical Imaging, Churchill Livingstone, (1982)

\item   Brian Cowan, Nuclear Magnetic Resonance and Relaxation, Cambridge University Press, (1997) 

\item G. B. Arfken, H. Weber, Mathematical Methods for Physicists, Academic Press, (2000) 

\end{enumerate}

\pagebreak
\noindent
{\Large Figure captions}

\medskip
\noindent
{\large Figure 1a}

\medskip
\noindent
This graph consists of 25 evenly spaced data points over the time interval, $-1 \leq t \leq 1$. It is a graph of the exponential function,
$s(t) = e^{-t}$

\medskip
\noindent
{\large Figure 1b}

\medskip
\noindent
This graph is the graph of the sinusoidal function that is generated from the odd component of the power series of the exponential 
function.

\medskip
\noindent
{\large Figure 1c}

\medskip
\noindent
This graph is the graph of the sinusoidal function that is generated from the even component of the power series of the exponential 
function.

\medskip
\noindent
{\large Figure 1d}

\medskip
\noindent
This graph is the graph of the high frequency harmonic function that is generated from the power series of the exponential function.
The angular frequency corresponding to $\Omega  + \alpha $, is 23.56 $s^{-1}$.

\medskip
\noindent
{\large Figure 2a}

\medskip
\noindent
This graph consists of 25 evenly spaced data points over the time interval, $-1 \leq t \leq 1$. It is a graph of a signal, s(t) consisting
of the exponential function, and a noise term, $n(t)$, $s(t) = e^{-t} + n(t)$. The noise is normally distributed with a mean of 0.0 
(Units), and a standard deviation of 0.01 (Units).

\medskip
\noindent
{\large Figure 2b}

\medskip
\noindent
This graph is the graph of the high frequency harmonic function that is generated from the power series of the signal s(t).
The angular frequency corresponding to $\Omega  + \alpha $, is 23.56 $s^{-1}$.

\pagebreak
\noindent
{\large Figure 3a}

\medskip
\noindent
This graph consists of 250 evenly spaced data points over the time interval, $-1 \leq t \leq 1$. It is a graph of a signal, s(t) 
consistingof the exponential function, and a noise term, $n(t)$, $s(t) = e^{-t} + n(t)$. The noise is normally distributed with a mean 
of 0.0 (Units), and a standard deviation of 0.05 (units).

\medskip
\noindent
{\large Figure 3b}

\medskip
\noindent
This graph is the graph of the high frequency harmonic function that is generated from the power series of the signal s(t).
The angular frequency corresponding to $\Omega  + \alpha $, is 23.56 $s^{-1}$.

\medskip
\noindent
{\large Figure 4a}

\medskip
\noindent
This graph consists of 250 evenly spaced data points over the time interval, $-1 \leq t \leq 1$. It is a graph of a signal, s(t) 
consistingof the exponential function, and a noise term, $n(t)$, $s(t) = e^{-t} + n(t)$. The noise is normally distributed with a mean 
of 0.0 (Units), and a standard deviation of 0.1 (units).

\medskip
\noindent
{\large Figure 4b}

\medskip
\noindent
This graph is the graph of the high frequency harmonic function that is generated from the power series of the signal s(t).
The angular frequency corresponding to $\Omega  + \alpha $, is 23.56 $s^{-1}$.

\medskip
\noindent
{\large Figure 5a}

\medskip
\noindent
This graph consists of 250 evenly spaced data points over the time interval, $-1 \leq t \leq 1$. It is a graph of a signal, s(t) 
consistingof the exponential function, and a noise term, $n(t)$, $s(t) = e^{-t} + n(t)$. The noise is normally distributed with a mean 
of 0.0 (Units), and a standard deviation of 0.5 (units).

\medskip
\noindent
{\large Figure 5b}

\medskip
\noindent
This graph is the graph of the high frequency harmonic function that is generated from the power series of the signal s(t).
The angular frequency corresponding to $\Omega  + \alpha $, is 23.56 $s^{-1}$.

\pagebreak
\begin{figure}[tp]
\includegraphics{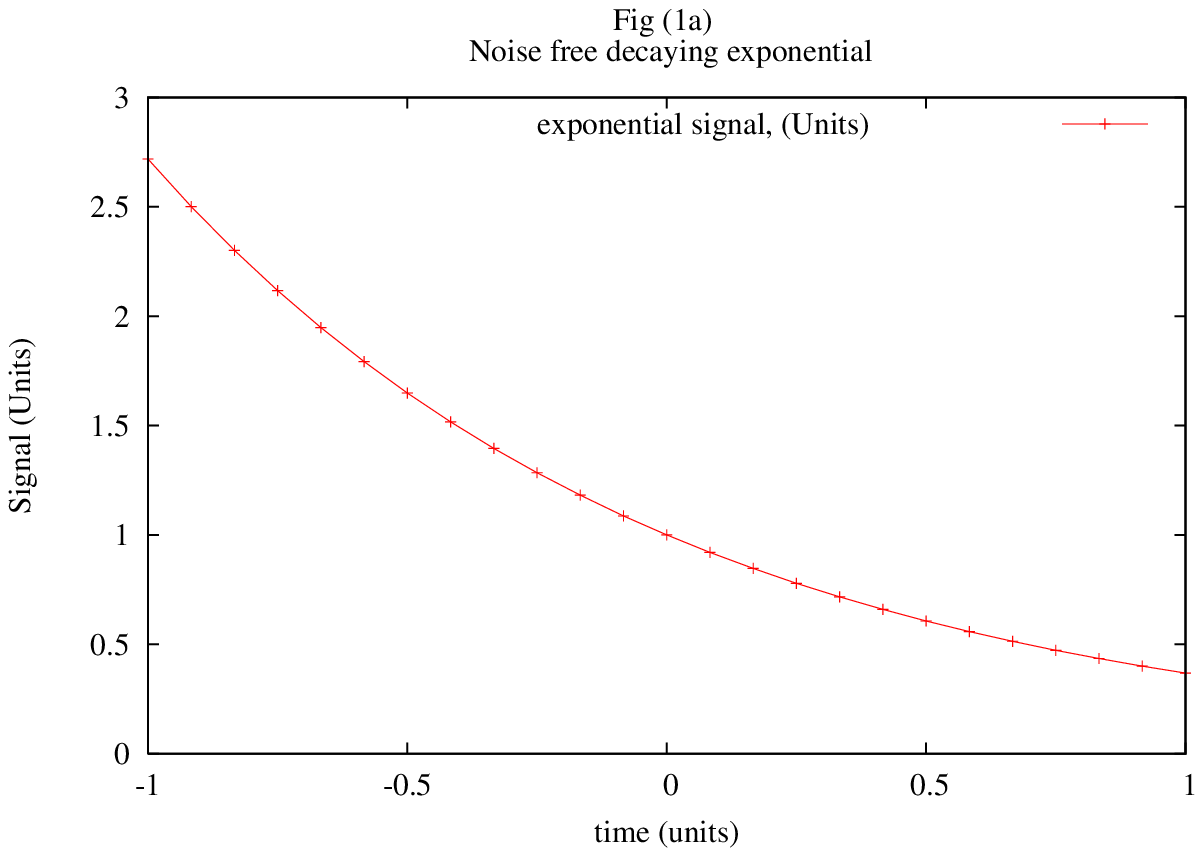}
\end{figure}

\begin{figure}[hp]
\includegraphics{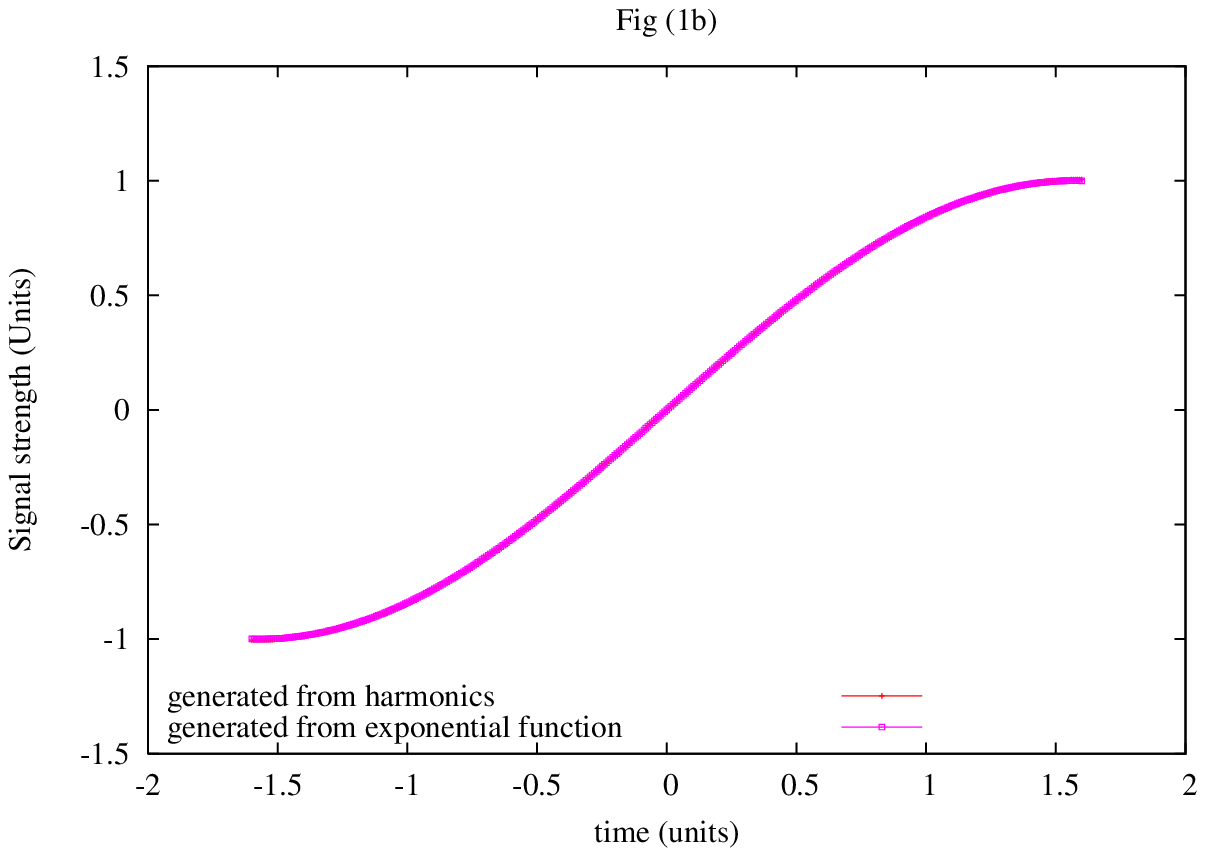}  
\end{figure}

\pagebreak
\begin{figure}[hp]
\includegraphics{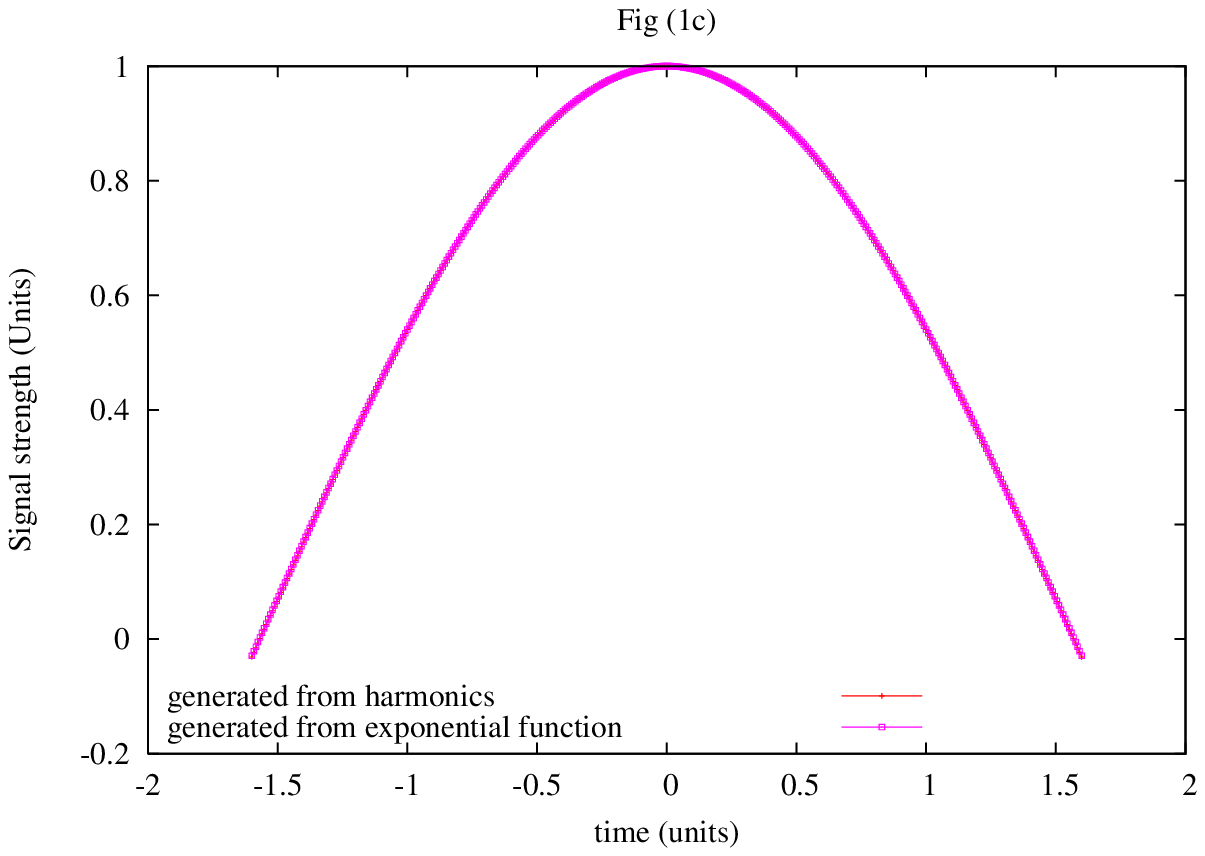}
\end{figure}

\begin{figure}[hp]
\includegraphics{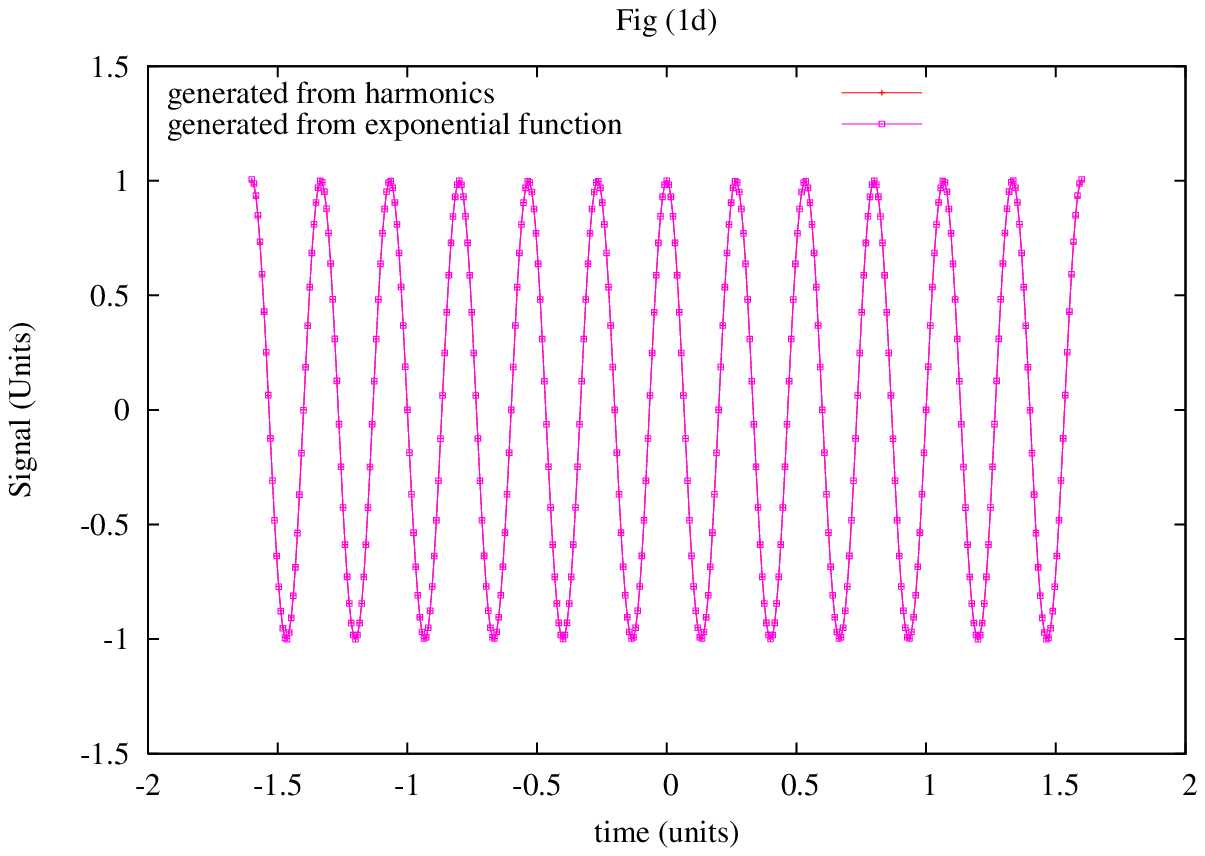}
\end{figure}

\pagebreak
\begin{figure}[hp]
\includegraphics{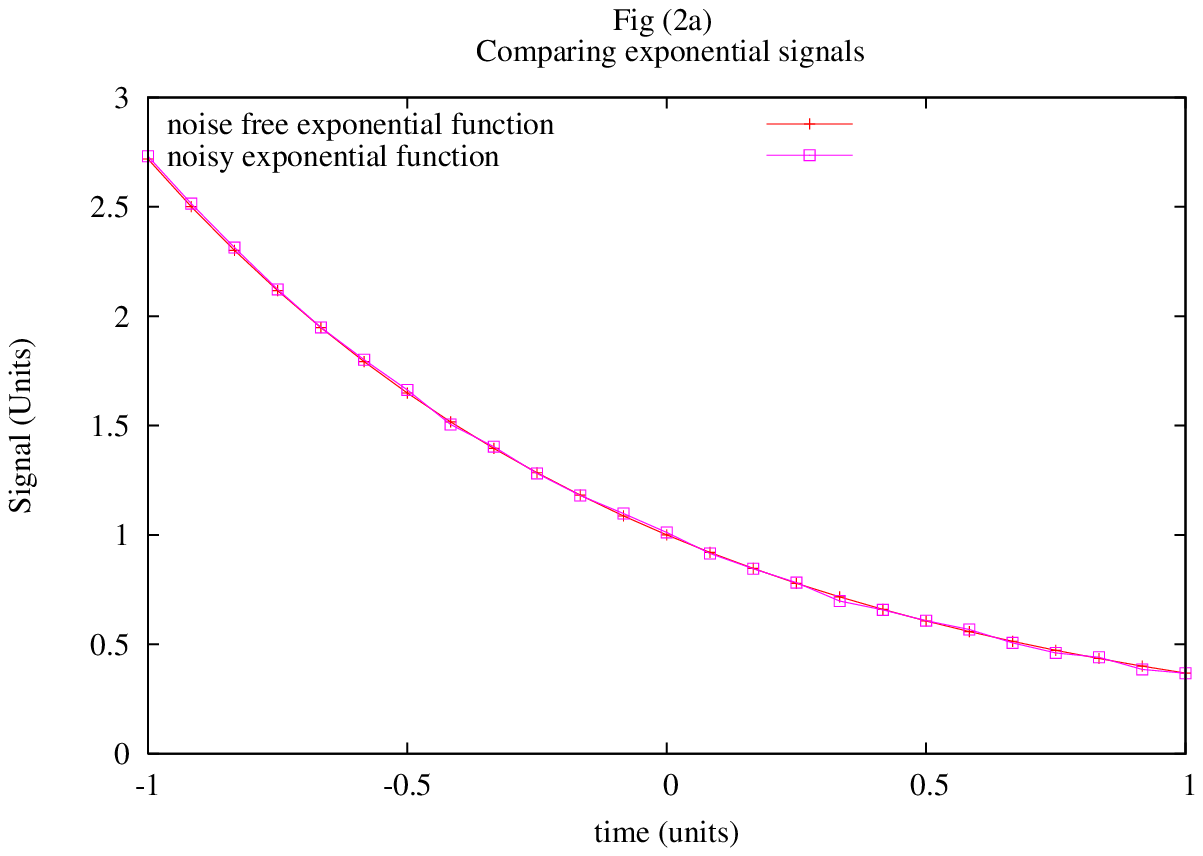}
\end{figure}

\begin{figure}[hp]
\includegraphics{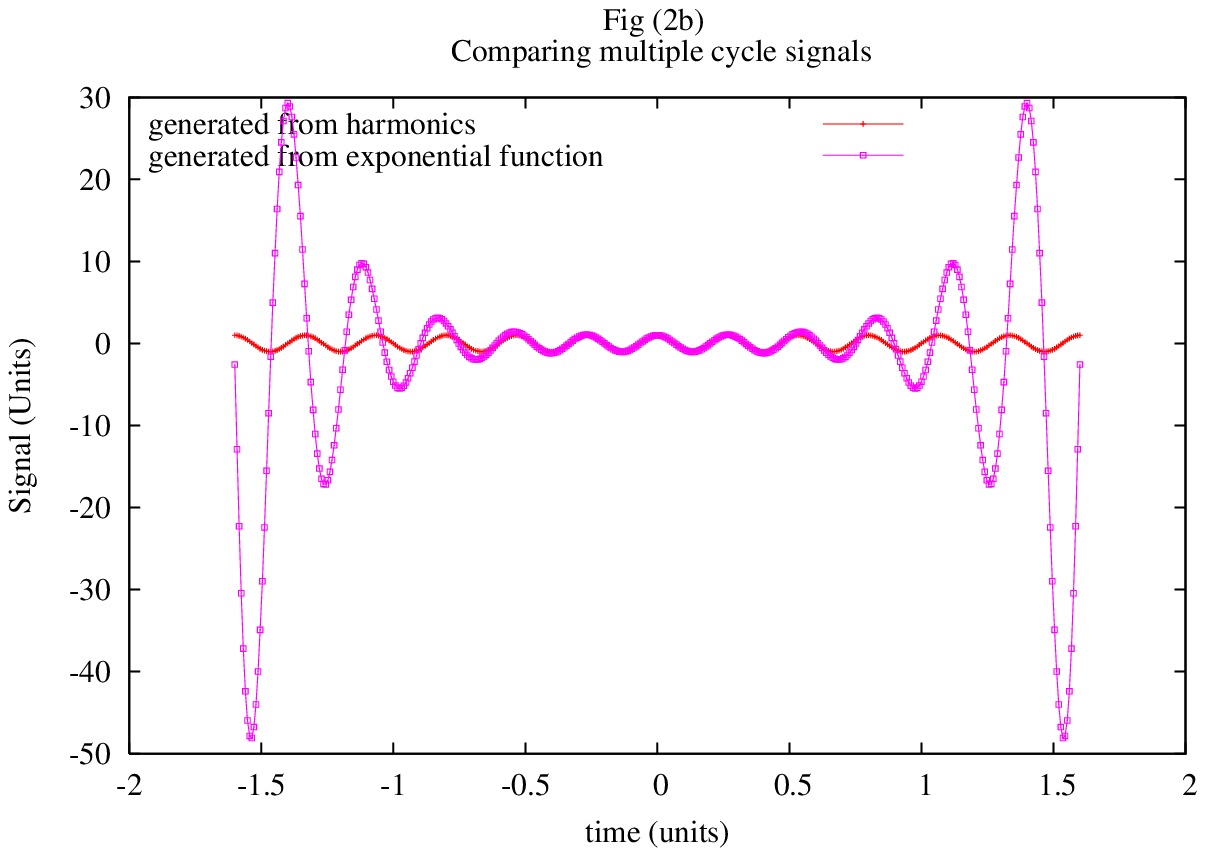}
\end{figure}

\pagebreak
\begin{figure}[hp]
\includegraphics{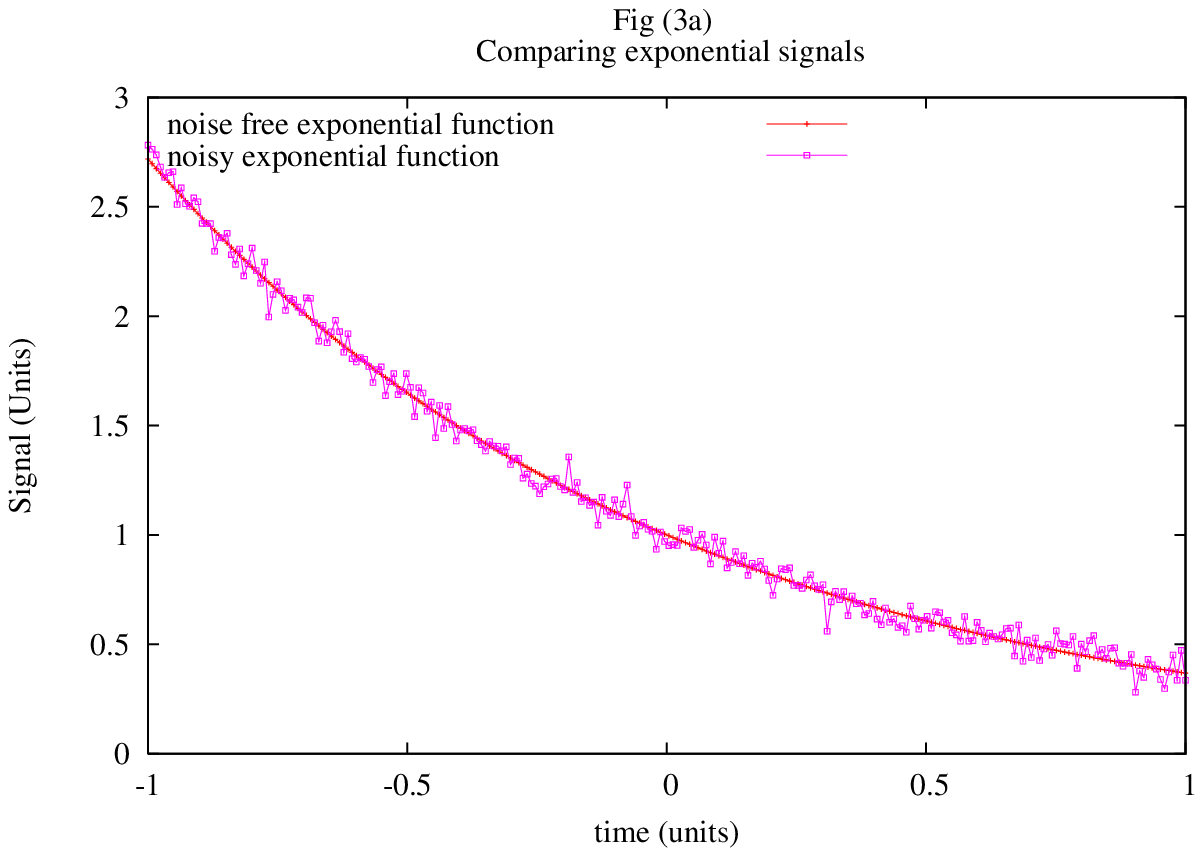}
\end{figure}

\begin{figure}[hp]
\includegraphics{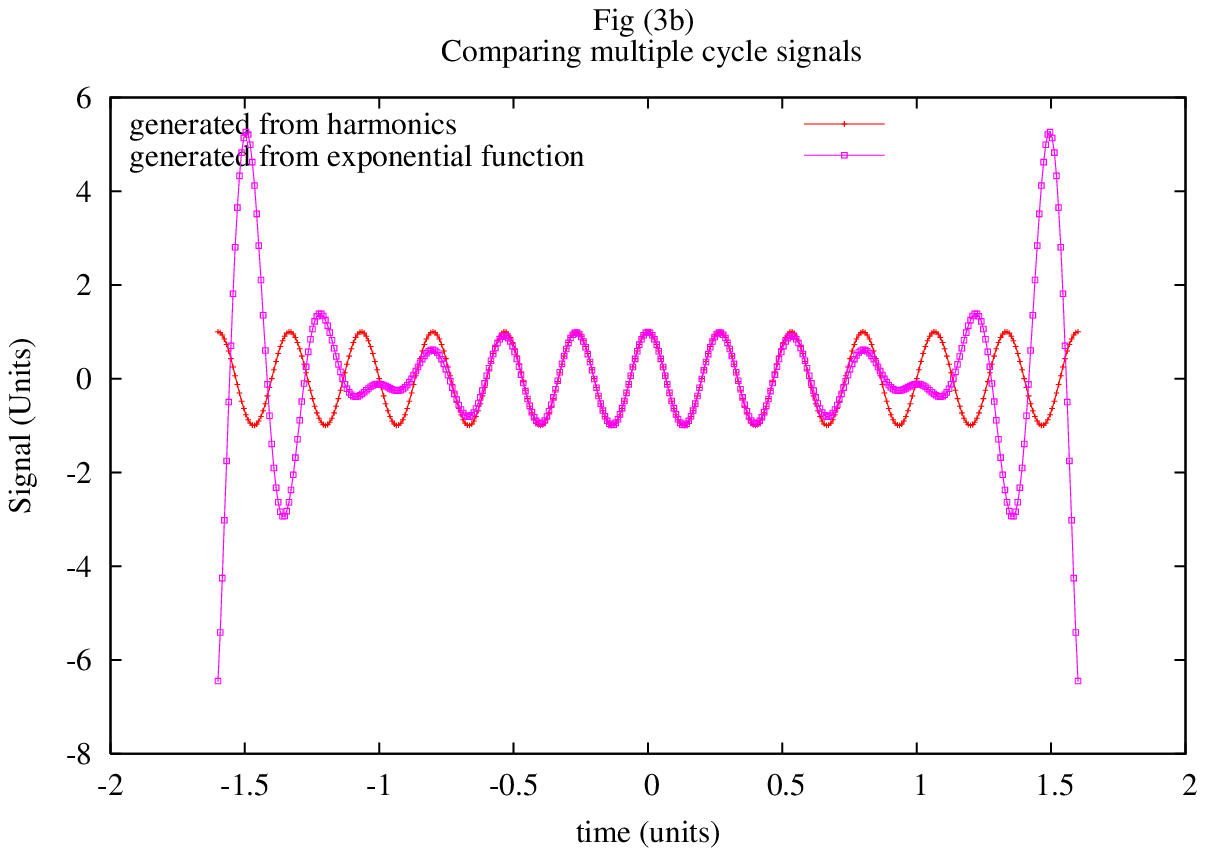}
\end{figure}

\pagebreak
\begin{figure}[hp]
\includegraphics{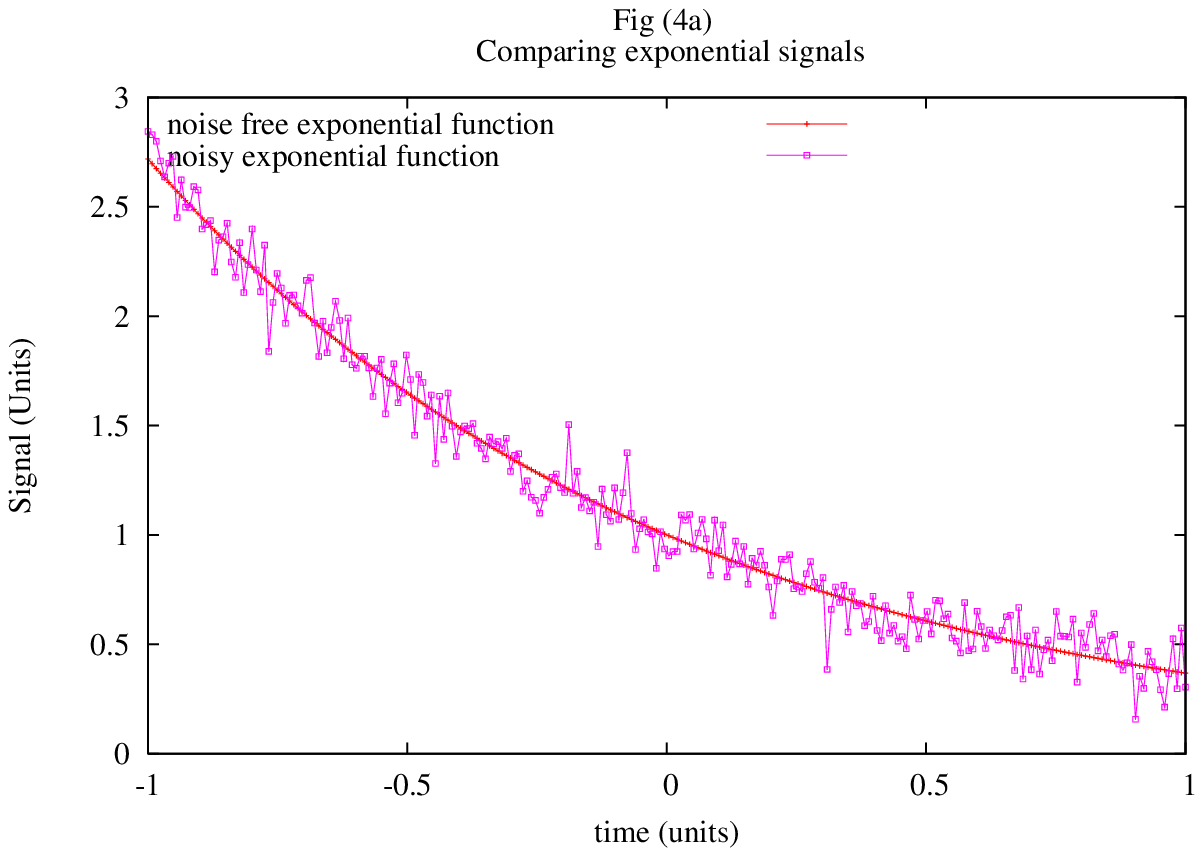}
\end{figure}

\begin{figure}[hp]
\includegraphics{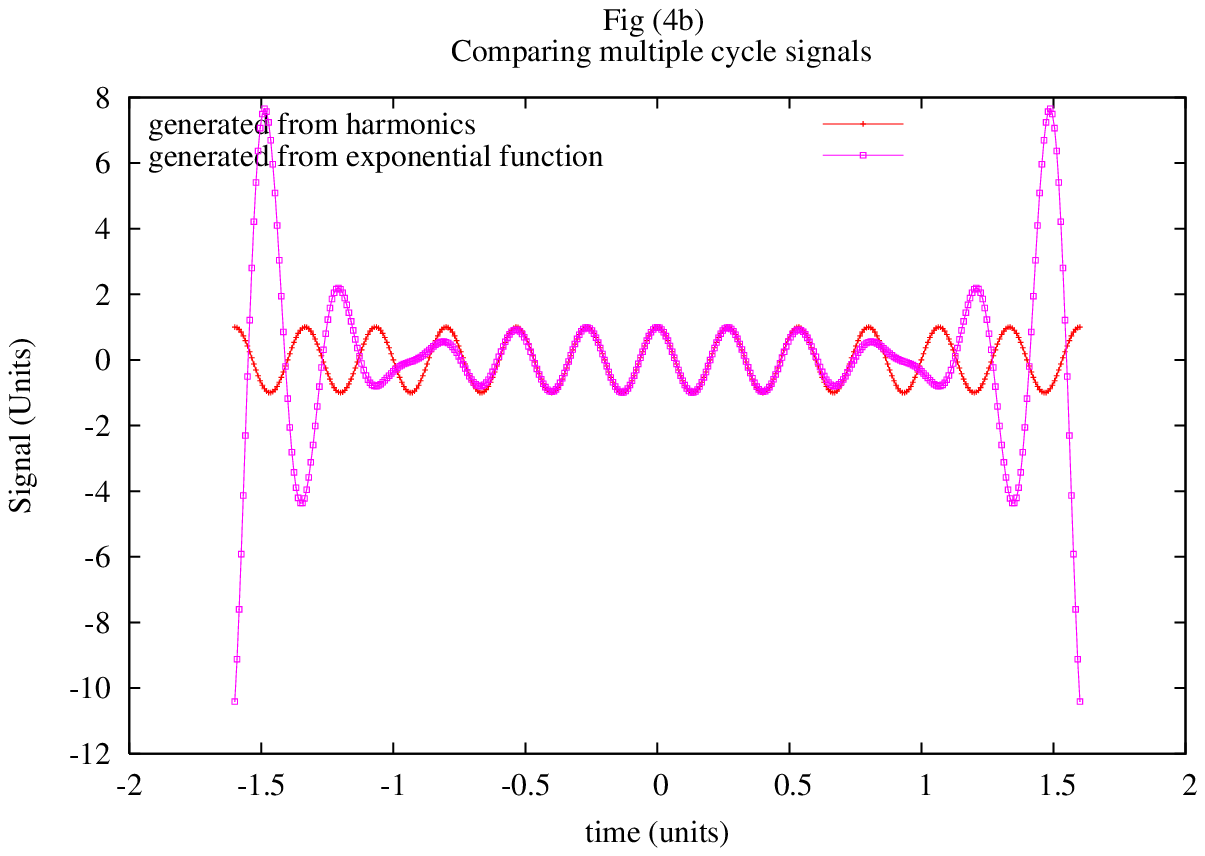}
\end{figure}

\pagebreak
\begin{figure}[hp]
\includegraphics{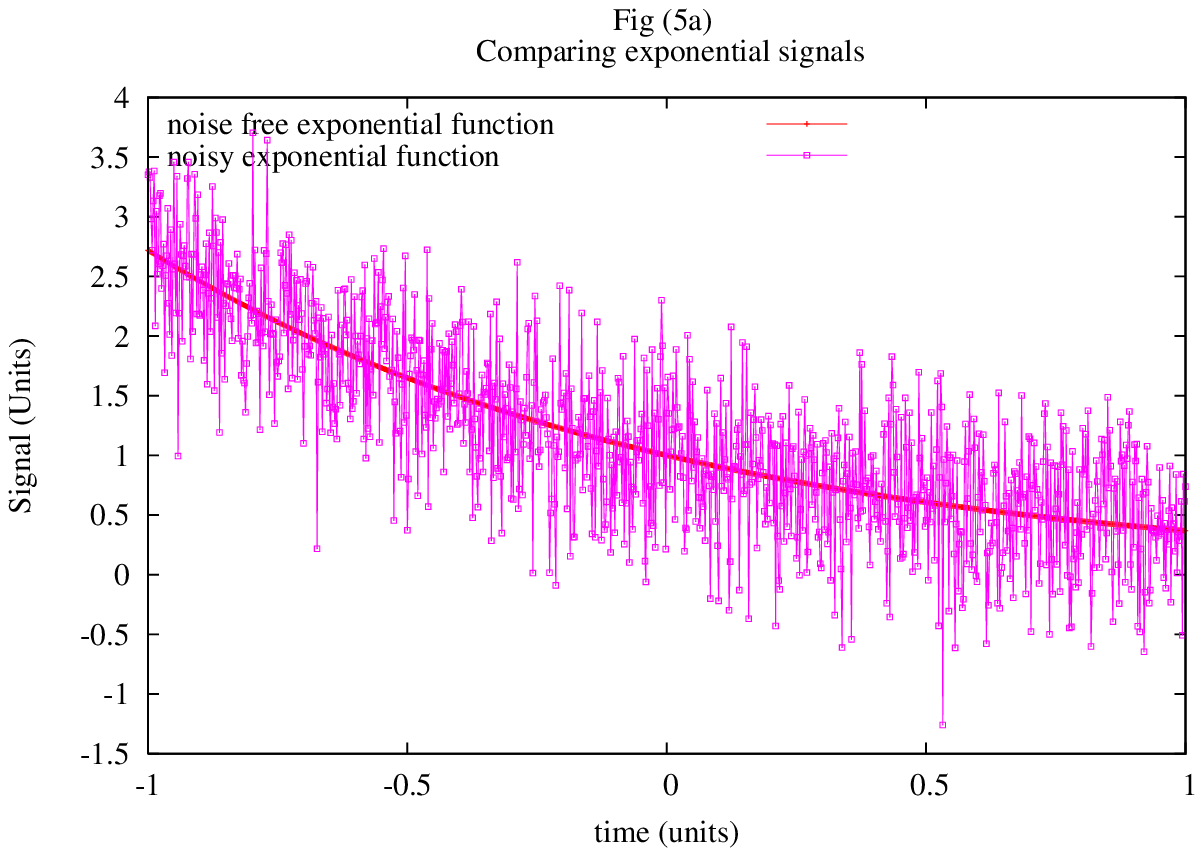}
\end{figure}

\begin{figure}[hp]
\includegraphics{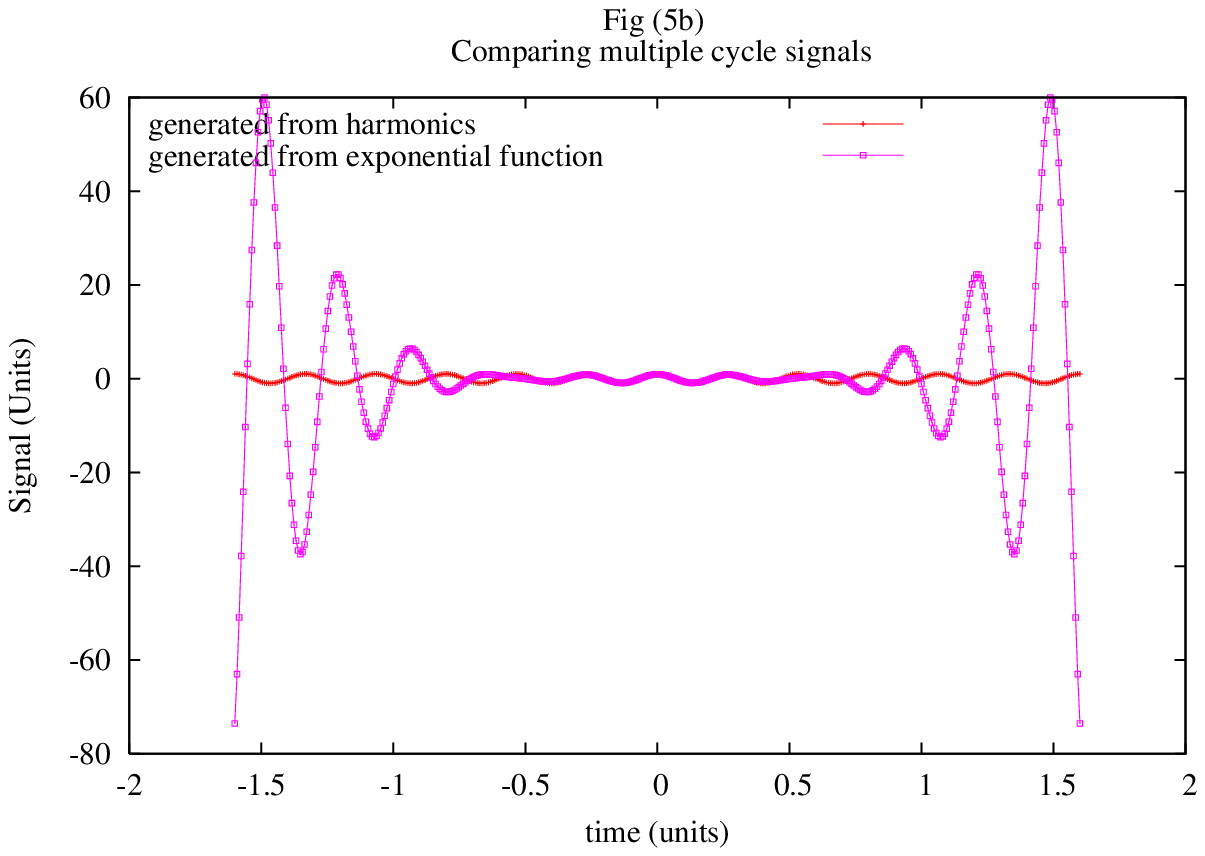}
\end{figure}

\end{document}